\documentclass{article}
\usepackage[switch]{lineno}

\usepackage[utf8]{inputenc}
\usepackage{balance}
\usepackage[dvips]{graphicx}
\usepackage{algorithmic}
\usepackage{algorithm}
\usepackage{listings}
\usepackage[OT4,T1]{fontenc}
\usepackage[cmex10]{amsmath}
\usepackage{url}
\usepackage{multirow}

\usepackage{pgfplots} 
\usepackage{tikz}
\usepackage{mathrsfs}
\usepackage{amssymb}
\usepackage{graphicx}
\usepackage{amsmath}
\usepackage{latexsym}
\usepackage{amsthm}
\usepackage{subfig}

\usepackage{algorithmic}

\usepackage{tabularx}
\usepackage{afterpage}
\usepackage{multirow}
\usepackage{float}
\newfloat{algorithm}{t}{lop}

\usepackage{url}

\usepackage{todonotes}

\newcommand{\footremember}[2]{%
    \footnote{#2}
    \newcounter{#1}
    \setcounter{#1}{\value{footnote}}%
}
\newcommand{\footrecall}[1]{%
    \footnotemark[\value{#1}]%
} 
\title{Optimization of Retrieval Algorithms on Large Scale Knowledge Graphs}

\author{%
	Jens Dörpinghaus\footnote{jens.doerpinghaus@scai.fraunhofer.de} \footremember{fh}{Fraunhofer Institute for Algorithms and Scientific Computing SCAI, Schloss Birlinghoven, Sankt Augustin, Germany}%
	\and Andreas Stefan\footrecall{fh}%
	}

\begin{document}
\maketitle              


\begin{abstract}
Knowledge graphs have been shown to play an important role in recent knowledge mining and discovery, for example in the field of life sciences or bioinformatics. Although a lot of research has been done on the field of query optimization, query transformation and of course in storing and retrieving large scale knowledge graphs the field of algorithmic optimization is still a major challenge and a vital factor in using graph databases. Few researchers have addressed the problem of optimizing algorithms on large scale labeled property graphs.
Here, we present two optimization approaches and compare them with a naive approach of directly querying the graph database. The aim of our work is to determine limiting factors of graph databases like Neo4j and we describe a novel solution to tackle these challenges. For this, we suggest a classification schema to differ between the complexity of a problem on a graph database.
We evaluate our optimization approaches on a test system containing a knowledge graph derived biomedical publication data enriched with text mining data. This dense graph has more than 71M nodes and 850M relationships. 
The results are very encouraging and -- depending on the problem -- we were able to show a speedup of a factor between 44 and 3839.
\end{abstract}

Although graph databases are a new field with constantly emerging technologies often missing common standards (like query languages) a lot of research has been done on the field of query optimization, query transformation and of course in storing and retrieving large scale knowledge graphs. While current state of the art systems often use RDF data models which are a collection of nested graphs and SPARQL queries the field is now driven by labeled property graphs to overcome their serious limitations. For example nodes and edges have no internal structure which does not allow complex queries like subgraph matchings or traversals and it is not possible to uniquely identify instances of relationships which have the same type, see \cite{desai2018issues}.

Here, we will present research on a more general topic related to large-scale optimization in parallel and distributed computational environments: Optimization of graph algorithms using queries to communicate with a graph database backend. We present two optimization approaches and compare them with a naive approach of directly querying the graph database. 

The topic of graph algorithms and their applications is widely studied in computer science and discrete mathematics. Using a graph database as data backend, graph algorithms rely on the robustness and velocity of the underlying system. This is  according to our knowledge a still unconsidered topic. We will focus on a particular graph database system (Neo4j) and consider the optimization of graph algorithms an dense large scale labeled property graphs with more then 71M nodes and 850M edges. They are based on biomedical knowledge graphs, see \cite{dorpinghaus2019knowledge}.

Communication with the database system might either be a complex query involving heuristics (like "give me all paths from node $a$ to $b$) or a simple query asking for a data set (like "give me all neighbors of node $a$) which are usually considered to take $\mathcal{O}(1)$ time. 

As a naive approach, we might expect that the runtime will not change using a graph database backend. If we want to find shortest paths between two nodes $a$ and $b$, we can rely on a build in function. We found, that for some nodes the database backend crashed due to insufficient memory. As a second try, we can use more simple queries. For example Dijkstra's algorithm is well known to have a time complexity of  $\mathcal{O}(m+n\cdot \log(n))$ given a graph $G=(V,E)$ with $|V|=n$ and $|E|=m$, see \cite{johnson1977efficient}. Here, we only need to retrieve the whole set of nodes and regularly the neighborhood of nodes and the weight of edges. Although these retrievals are considered to take $\mathcal{O}(1)$ time we have serious time problems to retrieve a dataset of 71M nodes using the Neo4j API. Using the graph database adds a factor based on a complex clew containing database efficiency, memory and computing power, connection speed and much more. 

This little example illustrates, that the usage of graph databases has serious algorithmic challenges not covered by computing complexity. The underlying challenges are related and not limited to query optimization, scaling and sharding technologies for databases and parallel algorithms. 
We will give an overview about this and other related work as a state of the art in the first section. 
After that, the second section  will give a brief overview about the background, infrastructure, data and research questions to solve. 
A novel, generic schema to categorize algorithms on graphs is presented in the third section. Here, we point at those candidates, where we need optimization approaches. 
The next section introduces three approaches to optimize graph queries. 
The fifth section presents an evaluation of these optimization strategies. After that, we will discuss the results and finish with conclusion and outlook.

\section{Related work}

It is obvious, that graph databases show different query times on different situations and there is a considerable amount of literature on that topic. For example an analyses of Neo4j and the performance of queries was done by \cite{huang2013research}. They show that there are difference in performance under different scenarios and they suggest query performance optimization for business applications. A review on storing big graphs in graph   databases  and their comparison is published by \cite{desai2018issues}. They conclude: "Graph  data  management  has  attracted  immense  research  attention though  it has  escaped  strong  foundations  of  designing  paradigms  for storage and retrieval. With growth and change in data with time, the need to  identify  patterns  and semantics  becomes difficult." We will present some recent related work which underlines this statement.

A lot of research has been done with respect to analyses and optimization of graph queries, especially with focus on Cypher and Neo4j. Hölsch and Grossniklaus \cite{holsch2016algebra} focus on an algebraic query transformation without the usage of a relational database system to process graph data. \cite{thakkar2017towards} conclude, that there is a very confusing situation, it is "an unforeseen race of developing new task specific graph systems, query languages and data models, such as property graphs, key-value, wide column, resource description framework (RDF)". They focus on Gremlin, which is a graph traversal language and machine to support multiple graph systems. They suggest a graph pattern matching for Gremlin queries supporting multiple graph data models. \cite{angles2019rdf} discuss issues of interoperability and optimization of queries between RDF and property graphs. They conclude, that more standards need to be developed. 
\cite{mennicke2019modal} questions about the general problems of knowledge graphs: "Although graph databases are conceived schema-less, additional knowledge about the data’s structure and/or semantics is beneficial in many graph database management tasks, from efficient storage, over query optimization, up to data integration." 
This is very plausible and we will highlight this possible pitfall in our work. 

A second topic in research is the technical optimization of the database. \cite{zhao2010graph} address the graph query problem on large networks by decomposing shortest paths around vertex neighborhood as basic indexing unit. This was found superior to GraphQL.
For Neo4j there are also several approaches. \cite{eymertoward} suggest a throughput optimization called in-graph batching which outperform standard Neo4j for large datasets. This is a similar approach to \cite{cheung2016sloth} who extended traditional lazy evaluation towards query batching while the application is executed. They noticed, that usually the communication, retrieval and storing of data is a crucial factor reducing the execution time of applications. This is exactly, what we noticed in our introduction example.  Other approaches have been proposed by \cite{mathew2018efficient} or \cite{cabrera2017scalable}.

Finally, a third way of optimization has to be mentioned. In the context of GIS graph databases \cite{wu2018research} try to optimize the heuristic for shortest paths. While also noticing the increasing time complexity for large graphs they tried to solve the problem using filters and adjusting the algorithms. These limitations were also found by \cite{yun2019tkg} while discussing the Frequent Subgraph Mining (FSM) task. Their novel TKG algorithm is also bound in size of the substructures analyzed in the graph database. One of the major drawbacks of these studies is that they focus on single problems in a very specific environment. 
There is still a considerable uncertainty with regard to algorithms and heuristics from a graph theoretical background when applying to graph databases.


\section{Background}

Using graph structures to house data has several advantages for knowledge extraction in life sciences and biological or medical research. Here, questions come from the field of exploring the mechanisms of living organisms and gaining a better understanding of underlying fundamental biological processes of life. 
In addition systems biology approaches, such as integrative knowledge graphs, are important as a holistic approach towards disease mechanism. In addition, pathway databases play an important role. As a basis, biomedical literature and text mining are used to build knowledge graphs, see \cite{dorpinghaus2019knowledge}. In addition relational data from domain specific languages like  BEL are widely applied to convert unstructured textual knowledge into a computable form. The BEL statements that form knowledge graphs are semantic triples that consist of concepts, functions and relationships \cite{fluck}. 
In addition, several databases and ontologies can implicitly form a knowledge graph. For example Gene Ontology, see \cite{GO} or DrugBank, see \cite{wishart2017drugbank} or \cite{khan2019consensus} cover a large amount of relations and references to which reference other fields. 

In \cite{huba} we collected 27 real world questions and queries in scientific projects to test the performance and output of the knowledge graph. We could show, that the performance of several queries was very poor and some of them even did not terminate. In order to identify limitations and understand the underlying problems, we carried on our work. 
The testing system is based an Neo4j and holds a dense large scale labeled property graph with more then 71M nodes and 850M edges. They are based on biomedical knowledge graphs as described in \cite{dorpinghaus2019knowledge}. 


\section{Classification of Problems}\label{sec:schema}

There seems to be no generally established procedure for categorizing graph-based queries. What we know about graph queries is largely based on six sources that categorize graph queries or describe them according to different criteria. The contents of this work and the results of the criteria are presented in this section.  

Author \cite{article} examines various theoretical classes of graph query languages with respect to the possible expressions and the complexity of evaluating queries. However, the study is not based on the property graph model, but on a simpler model with a finite directed graph with edge labels. 
In their analysis they show that for current graph databases, including Neo4j, there is a lack of a language with clear syntax and semantics.
They claim that this is a difficulty to evaluate the expressiveness and computational effort of possible queries.  

 
(1) \cite{6313676} describe graph queries that are considered relevant on the basis of the author's literature research and can be divided into the following four categories:
\begin{itemize}
\item{Adjacency queries}

Adjacency queries check whether two nodes are connected or in the \textit{k neighborhood} of each other.
\item{Reachability queries}

Accessibility queries check whether a node can be reached via a fixed-length path or via a simple regular path and which is the shortest path between the nodes. 
\item{Pattern Matching queries}

Pattern matching consists of finding all subgraphs of a graph that are isomorphic to a pattern graph.
\item{Summarization queries}

These types of queries are based on functions that allow the results of the queries to be summarized, usually returning a single value. 
These include functions such as average, number, maximum, etc.
They also include functions for calculating properties of the graph and its elements such as the degree of a node, the minimum, maximum and average degrees in the graph, the length of a path, the distance between two nodes, the diameter of the graph, etc.
\end{itemize}

(2) \cite{Angles:2017:FMQ:3145473.3104031} divide graph queries into two basic functions: \textit{Graph Patterns}, where a pattern structured as a graph is searched in the database, and \textit{Graph Navigation}, which should find paths of any length. 
The graph pattern queries can be further restricted by projection, union and difference. The result of a Graph Pattern query is a set of all mappings of variables from the query to constants in the database.


The simplest query in the class of Graph Navigation Queries is wether a certain path exists in the graph. This can be extended by additional restrictions, for example, by allowing only certain edge labels. To do this, a path query can be described in general terms as $P=x\longrightarrow{\alpha}y$, where $\alpha$ specifies the restrictions. The endpoints $x$ and $y$ can be variables or specific nodes. 
The best known formalism for representing $\alpha$ is \textit{regular expressions}. Regular expressions allow the concatenation of paths and the application of a union or disjunction of paths. 
Path queries specified with regular expressions are commonly referred to as \textit{Regular Path Queries (RPQ)}. 
\cite{Angles:2017:FMQ:3145473.3104031} provide information on the complexity of evaluating RPQs to determine whether a path exists. However, the complexity information for RPQs cannot simply be applied to Cypher. 
In addition, they show that everal open questions regarding complexity or the graph query language Cypher remain. In contrast to SPARQL, the semantics and complexity of Cypher has not yet been investigated due to the lack of theoretical formalization, see \cite{Angles:2017:FMQ:3145473.3104031} and \cite{Wood:2012:QLG:2206869.2206879}.

(3) \cite{Wood:2012:QLG:2206869.2206879} describe different classes of queries for several graph query languages, as well as several core functionalities supported by the graph query languages. They also discuss the expressiveness and complexity of query evaluation. Unfortunately, Cypher is not described as a graph query language. The author divides the queries into the following categories:
\begin{itemize}
\item{\emph{CQ} (conjunctive query)}

A sample query of this type looks for documents that have both the \textit{PublicationType} \texttt{Journal Article} and \texttt{Review}. 
\item{\emph{RPQ} (regular path query)}

A search is made for a node pair $(x,y)$ so that a path exists between $x$ and $y$, with the sequence of edge labels following a given pattern. The given pattern is described by a regular expression. 
\item{\emph{CRPQ} (conjunctive regular path queries)}

\emph{CQ}s and \emph{RPQ}s can be combined to form the class \emph{CRPQ}. According to the author, this class serves as a basis for several graph query languages. However, this class is not sufficient for problems where relationships between paths need to be specified. 
\item{\emph{ECRPQ} (extended conjunctive regular path query}

This class extends the \emph{CRPQ}s by the possibility to specify path variables or to allow paths as output of a query. 
\end{itemize}

\begin{figure*}[t] 
	\centering
	\includegraphics[width=0.95\textwidth]{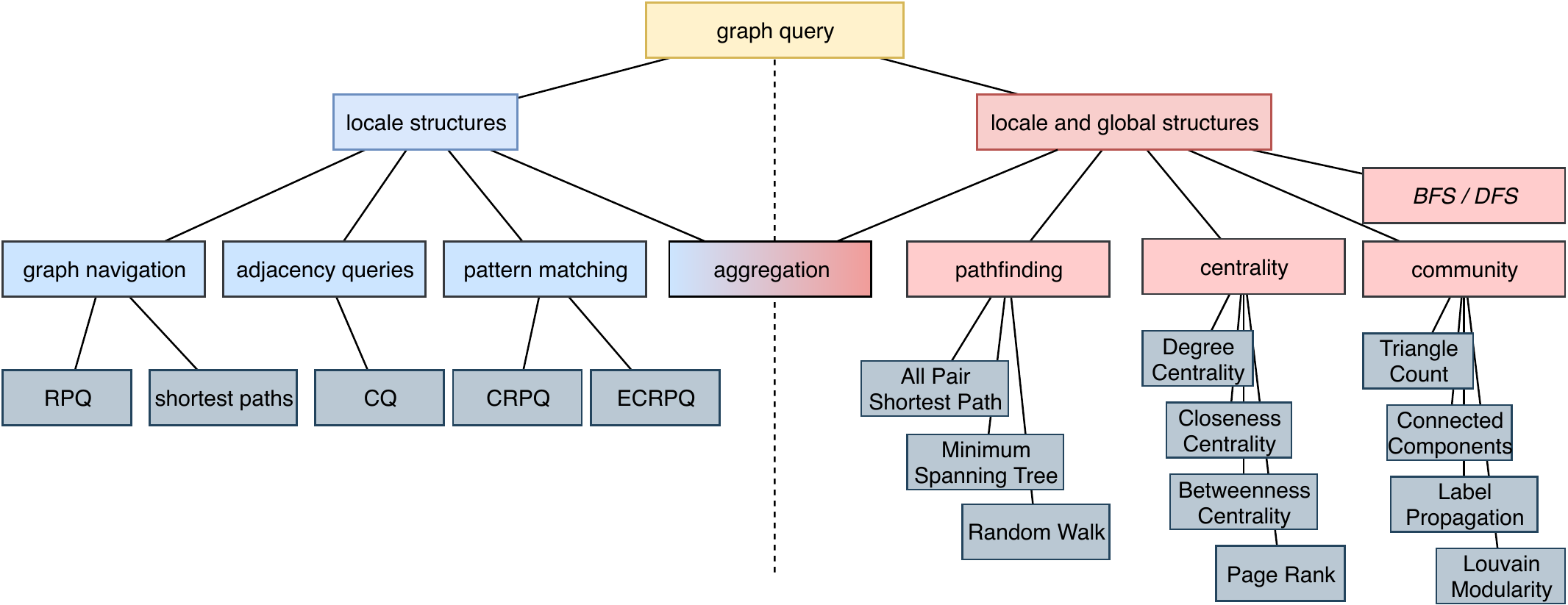}
	\caption{An overview of the categories for graph queries unified from literature sources. These categories give a first overwiev and a categorization scheme for graph queries and their complexity.}
	\label{img:visualization}
\end{figure*} 

In addition \cite{Wood:2012:QLG:2206869.2206879} examine functionalities of graph query languages. They are divided into the following categories:
\begin{itemize}
\item{Subgraph Matching}

It searches for subgraphs in a graph. This is a \emph{CQ}.
\item{Find connected nodes by path} 

Determining accessibility between nodes in a graph is a graph query that is supported in many graph query languages. The \emph{RPQ}s class includes queries that return all node pairs from a graph that are connected by a path that matches a regular expression.  
\item{Compare and return paths} 

It specifies relationships between paths and searches for paths that connect two nodes to find connections in linked data.
By providing these two functions, the class of extended \emph{CRPQ}s (\emph{ECRPQ}s) is created.
\item{Aggregation}

Determining different properties of graphs requires a calculation that goes beyond matching and finding paths. Such properties are for example the determination of node degrees. 
\end{itemize}

(4) Both \cite{Pokorny2018} and \cite{10.1007/978-3-319-24369-6_5} consider queries with property graphs and name among others Cypher and Gremlin as important graph query languages. These sources name these categories of graph queries:
\begin{itemize}
\item{k-hop Queries}

According to the authors, these queries are most common in practice. They include queries such as \textit{find node}, \textit{find the node's neighbors (1-hop query)}, \textit{find edges in multiple hops}, and \textit{get attribute values}.
\item{subgraph and supergraph queries}
\item{width search / depth search}
\item{Seeking and shortcuts}
\item{Search for strongly connected components}
\item{Regular Path Queries}
\end{itemize}


(5) In \cite{Needham2019}, queries and graph algorithms are described and subdivided according to different properties. On the one hand, the authors subdivide the queries according to \textit{graph pattern}-based queries for local analysis of the data and according to \textit{graph algorithms}, which often analyze globally and iteratively.
 Local queries only look at a specific section of the graph like a start node and the surrounding subgraph. This type of query is often used for transactions and pattern-based queries. Graph algorithms typically search for global structures. The algorithm takes the entire graph as input and returns an enriched graph or an aggregated value. 

The authors divide different graph algorithms into the three categories \textit{Pathfinding}, \textit{Centrality} and \textit{Community Detection}. The book describes several graph algorithms and assigns them to the categories:

\begin{itemize}
\item pathfinding
	\begin{itemize}
	\item Shortest Paths
	\item All Pair Shortest Path
	\item Minimum Spanning Tree
	\item random walk
	\end{itemize}
\item Centrality
	\begin{itemize}
	\item Degree Centrality
	\item Closeness Centrality
	\item Betweenness Centrality
	\item page rank
	\end{itemize}
\item Community Detection
	\begin{itemize}
	\item Triangle Count
	\item (Strongly) Connected Components
	\item Label Propagation 
	\item Louvain Modularity
	\end{itemize}
\end{itemize}

\subsection{New criteria}
In order to categorize graph queries, we introduce new criteria, which were found relevant for the use case evaluated for our knowledge graph:
\begin{itemize}
\item{Accessing attributes}

How many attributes must be considered when executing the query? Accessing attributes requires reading an additional file and therefore requires more processing power and access time. In section \ref{chap:eval} we will proof, that data stored in attributes will significantly slow down queries.
\item{Data type of attributes}

What data types are accessed in queries? We expect, that this also influences the runtime. 

\item{Node and edge types to be considered}

Which node and edge types must be considered in the query? Is it only a small subset or is the majority of the types required? Is it possible to decide for all queries whether and which node and edge types can be exported as subgraphs?

\item{Entry point}

Does the query rely on a unique node specified for the query (e.g. as a starting point for the search), or 
is there a general search for pattern between nodes?  
\end{itemize} 

Various approaches have been proposed in literature, but we can examine connections and a hierarchy. In the next step, we merge and cluster these approaches in order to create a categorization scheme for graph queries. This is shown in figure \ref{img:visualization}.

The schema divides the categories for graph queries into \textit{local structures} and \textit{local \& global} structures (according to literature source (5)). The second structure category is called \textit{local \& global}, because some of the graph algorithms can act locally by specifying a start node or a subgraph. 
Furthermore, some categories, such as \emph{CRPQ} or \emph{ECRPQ}, were identified as subcategory of other categories. This is illustrated by the hierarchical structure of the categorization scheme.  
The category \textit{Aggregation} belongs to graph queries that search for both local and global structures. For example, the category \textit{Aggregation} can include questions such as "What is the degree of node A?" or "What is the average of the graph?", the former referring to local and the latter to global structures.

\section{Method}

\begin{figure*}[t] 
	\centering
	\includegraphics[width=0.95\textwidth]{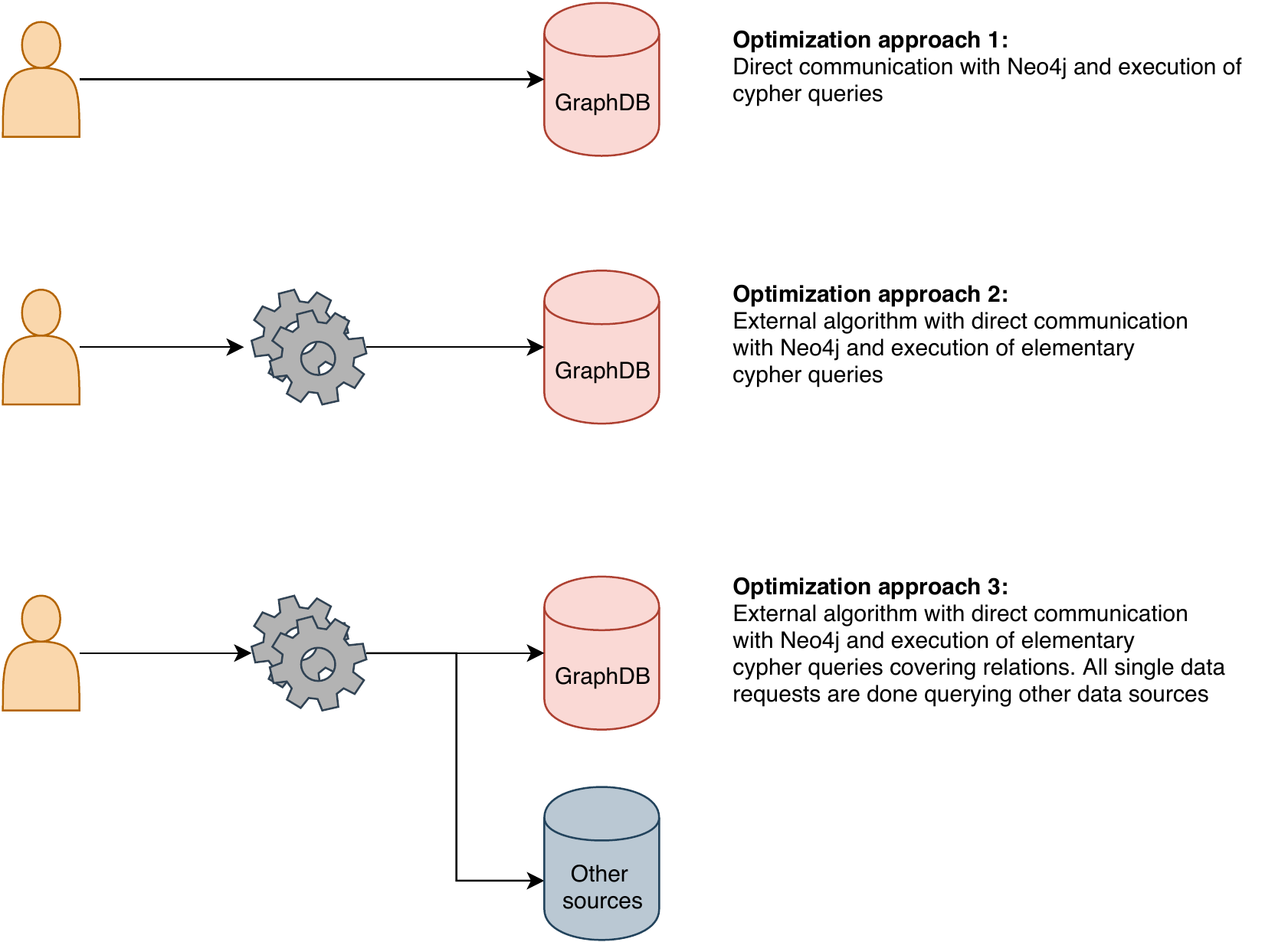}
	\caption{An overview of the optimization approaches discussed in this paper. The first approach contains the basic Cypher query, the second approach transfers the algorithm to a different system. The third approach relies on a polyglot persistence architecture and excludes all time-consuming queries that can be answered by a key-value store.}
	\label{img:opti}
\end{figure*}

Here, we propose a multi-step optimization approach towards graph queries. Usually, graph queries are executed using a Cypher query. Here, the application or the user directly communicates with the graph database. To optimize this, we suggest that an external algorithm communicates with the graph database and executes only elementary queries. With this, the queries are limited to typical questions like neighborhood, paths and relations. Since all trivial requests (like "give me this node") can usually be handled by common relational or special purpose databases, we suggest a third optimization approach, if necessary. Here, a polyglot persistence approach uses other data sources to execute trivial queries. See figure \ref{img:opti} for an illustration. 

\subsection{Pathfinding}

In \cite{huba} we introduced a large set of queries and categorized them according to the schema discussed in section \ref{sec:schema}. We will start with those problems using in general both locale as well as global structures in the graph. A problem with a very poor performance was graph navigation and pathfinding. These include Regular Path Queries (RPQ, see \cite{Angles:2017:FMQ:3145473.3104031}) (problems 2,11,14,16,17,19,21) and finding shortest paths (problems 4,12). Since the problems of retrieving a single or all shortest paths are quite similar, we will discuss both of them here.

Queries 4 and 12 are both a typical shortest path problem: \emph{What is the shortest way between \{Entity1\} and \{Entity2\} and what is on that way?} and \emph{How far apart are \{document1\} and \{document2\}?} Thus both problems can be solved using Cypher:

\noindent(Q4) \texttt{match (entity1:Entity \{preferredLabel: "axonal transport"\}), (entity2:Entity \{preferredLabel: "LRP3"\}) call algo.shortestPath.stream(entity1, entity2) yield nodeId return algo.asNode(nodeId)}

\noindent(Q12) \texttt{match (doc1:Document \{documentID: "PMID:16160056"\}), (doc2:Document \{documentID: "PMID:16160050"\}) call algo.shortestPath.stream(doc1,doc2) yield nodeId return algo.asNode(nodeId)}

Both queries rely on the function \texttt{shortestPath} available in Neo4j. Both Bellman--Ford and Dijkstra's algorithm are known to solve this problem for weighted graphs. For unweighted graphs a modified  Breadth-first search will solve this issue in $\mathcal{O}(E + V)$ \cite{aziz2010algorithms}. Other algorithms like Dijkstra's should be faster, for example using binary heaps the time complexity is  $\mathcal{O}(m+n\cdot \log(n))$ given a graph $G=(V,E)$ with $|V|=n$ and $|E|=m$, see \cite{johnson1977efficient}. According to Neo4j documentation, the build in function \texttt{shortestPath} uses Dijkstra's algorithm\footnote{See \url{https://neo4j.com/docs/graph-algorithms/current/labs-algorithms/shortest-path/}.}.

\begin{algorithm}[h!]
\begin{algorithmic}[2]
\caption{\textsc{Graph-BFS}} \label{alg:bfs}
\REQUIRE two nodes $s,e\in V$ 
\ENSURE shortest path $p=[s, ... , e]$
\STATE $Q$ = []
\STATE $discovered = [s]$
\STATE $Parent = {}$
\STATE $Q.append (s)$
\WHILE {$len(Q)>0$}
\STATE $v = Q.pop(0)$
\IF{$getNode(v)==e$}
\STATE $x = v$
\STATE $path = [ v ]$
\WHILE{$Parent[x] != s$}
\STATE $x = Parent[x]$
\STATE $path.append(x)$
\ENDWHILE
\STATE $x = Parent[x]$
\STATE $path.append(x)$
\STATE $return path$
\ENDIF
\STATE $N = getNeighbours(v)$ \label{line:neigh}
\FOR{$w in N$}
\IF{$w$ not in $discovered$}
\STATE $discovered.append (w)$
\STATE $Parent[w] = v$
\STATE $Q.append (w)$
\ENDIF
\ENDFOR
\ENDWHILE
\RETURN $d$ with max ($pd$)
\end{algorithmic}
\end{algorithm}
                
With algorithm \ref{alg:bfs} we suggest a BFS-approach to tackle the shortest-path problems. Given both a starting node $s$ and an ending node $e$, the only communication with the graph database is done in line \ref{line:neigh}. Here, the neighborhood of a node is retrieved.  

This algorithm implements the optimization approach 2. Since no other data sources are needed, optimization approach 3 will not improve this query. 

\subsection{CRPQ}

Several questions introduced in \cite{huba} are conjunctive regular path queries (CRPQ, see \cite{Wood:2012:QLG:2206869.2206879}). These are pattern matching problems using locale structures within the graph. Some of them are quite simple. For example query 15 -- \emph{How many sources are there for the statements of a contradictory BEL statement?} -- can be easily translated into Cypher: 

\noindent(Q15) \texttt{match (e1:Entity) -[r1:hasRelation \{function:"increases"\}]-> (e2:Entity),  (e1) -[r2:hasRelation \{function:"decreases"\}] -> (e2) return distinct e1.preferredLabel, e2.preferredLabel, count(r1) as `increases`, count(r2) as `decreases` order by count(r1) desc}

This query matches two contradicting relations, their numbers and returns a decreasing sorted list. More complex is the example query 1: \emph{Which author was the first to state that \{Entity1\} has an enhancing effect on \{Entity2\}?} A Cypher query solving this uses several node attributes, for example the publication date to sort the result set:

\noindent{(Q1) \texttt{match (n:Entity {preferredLabel: "APP"}) -[r:hasRelation {function: "increases"}]-> (m:Entity {preferredLabel: "gamma Secretase Complex"}), (doc:Document {documentID: r.context}) <-[r2:isAuthor]- (author:Author) return doc, author order by doc.publicationDate limit 1'}}

As a first optimization approach denoted by opt1 we exclude the sorting functions from the queries and do this manually. This leads to the following two queries:

\noindent{(Q1-1) \texttt{match (n:Entity {preferredLabel: "APP"})-[r:hasRelation {function: "increases"}]->(m:Entity {preferredLabel: "gamma Secretase Complex"}) return n,r,m}}

\noindent{(Q15-1) \texttt{match (e1:Entity) -[r1:hasRelation {function:"increases"}]-> (e2:Entity),  (e1) -[r2:hasRelation {function:"decreases"}]-> (e2) return distinct e1.preferredLabel, e2.preferredLabel}}

The algorithm for query 1 can be found in \ref{alg:q1o1}, the algorithm for query 15 in \ref{alg:q15o1}. As we can see, query 1 is more complex, since it includes the retrieval of node attributes, the publication data. Both algorithms include the sorting of lists. 

\begin{algorithm}[h!]
\begin{algorithmic}[2]
\caption{\textsc{Query1-opt1}} \label{alg:q1o1}
\REQUIRE Documents $D=\{d_{1},...,d_{n}\}$ obtained from query (Q1-1)
\ENSURE Document $d$
\STATE $pd = []$
\FOR {every $d \in D$}
\STATE $pd$.add ($d$,$d$.publicationdate)\label{line:pd}
\ENDFOR
\RETURN $d$ with max ($pd$)
\end{algorithmic}
\end{algorithm}

\begin{algorithm}[h!]
\begin{algorithmic}[2]
\caption{\textsc{Query15-opt1}} \label{alg:q15o1}
\REQUIRE Data points $T=\{t_1, ..., t_n\}$ with $t_i=\{e1_i,e2_i, inc_i, dec_i\}$ obtained from query (Q15-1)
\ENSURE Sorted data points $T$
\RETURN sort($T$)
\end{algorithmic}
\end{algorithm}

The second optimization approach can only be applied to query 1. Here, we try to retrieve the node attributes from a dedicated information system. This is related to the polyglot persistence approach introduced in \cite{huba}. Here, we suggest to retrieve this value direct from the SCAIView API. 

\noindent{(Q1-2) \texttt{match (n:Entity {preferredLabel: "APP"})-[r:hasRelation {function: "increases"}]->(m:Entity {preferredLabel: "gamma Secretase Complex"}) return n,r,m}}

Here, algorithm \textsc{Query1-opt2} will use a different function to add the publicationdate in line \ref{line:pd}.


\section{Evaluation}\label{chap:eval}

We evaluate our optimization approaches on a test system containing a knowledge graph derived biomedical publication data enriched with text mining data and domain specific language data using BEL, see \cite{huba}. This dense graph has more than 71M nodes and 850M relationships. 

The testing system run Neo4j Community 3.5.8. on a server with 16 Intel Xeon CPUs with 3GHz and 128GB main memory. We applied several approaches described in the chapter "Performance" in the Neo4j Operations Manual\footnote{See \url{https://neo4j.com/docs/operations-manual/current/performance/}.}.  

\subsection{Pathfinding}

\begin{figure*}[t] 
	\centering
	\includegraphics[width=0.95\textwidth]{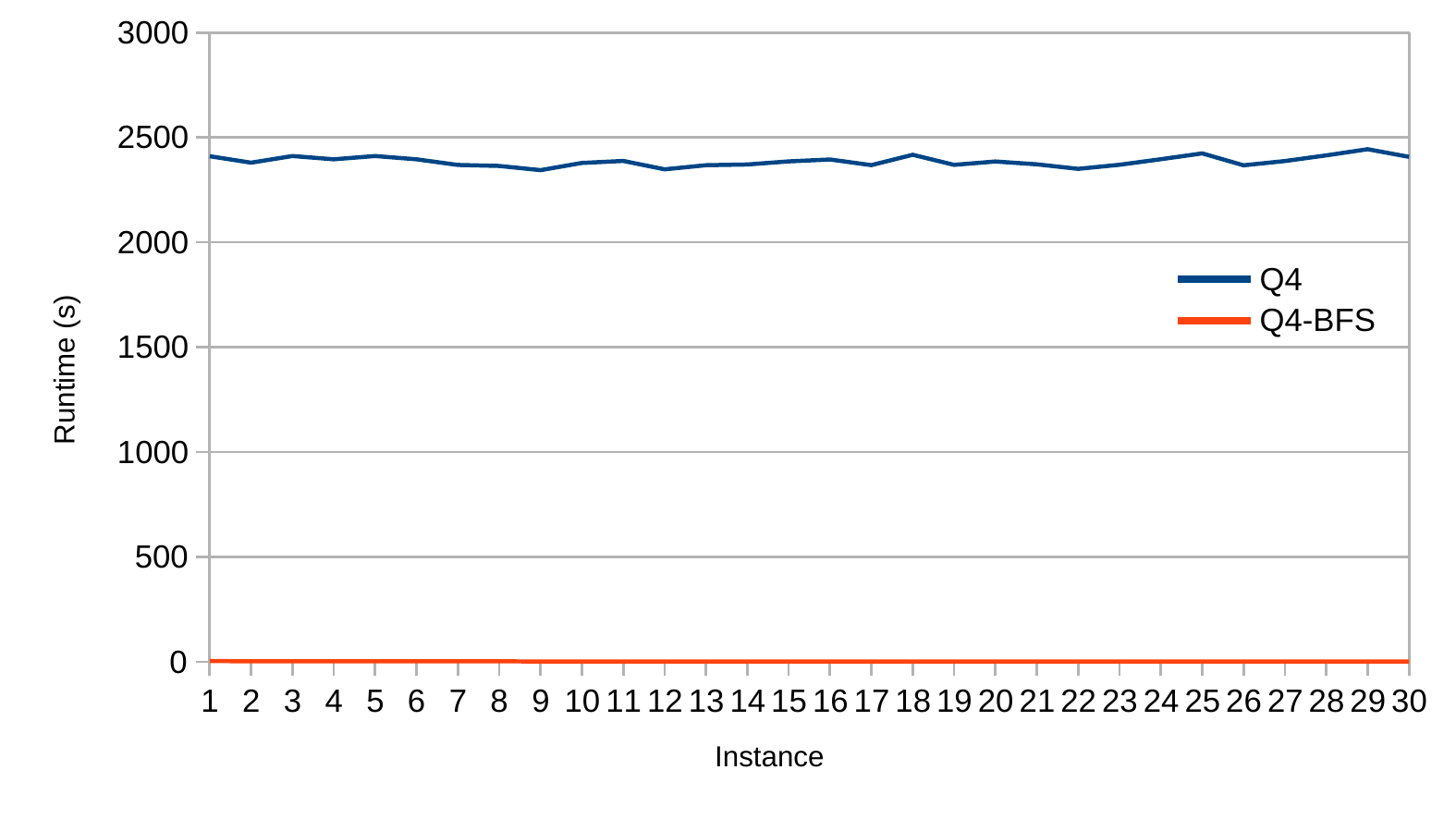}
	\caption{Results for query 4 \textsc{Query4} (average runtime 2390.44 seconds) and the optimization approach 1 \textsc{Graph-BFS} (average runtime 1.65 seconds). The speedup factor is 1453.}
	\label{img:q4}
\end{figure*} 

\begin{figure*}[t] 
	\centering
	\includegraphics[width=0.95\textwidth]{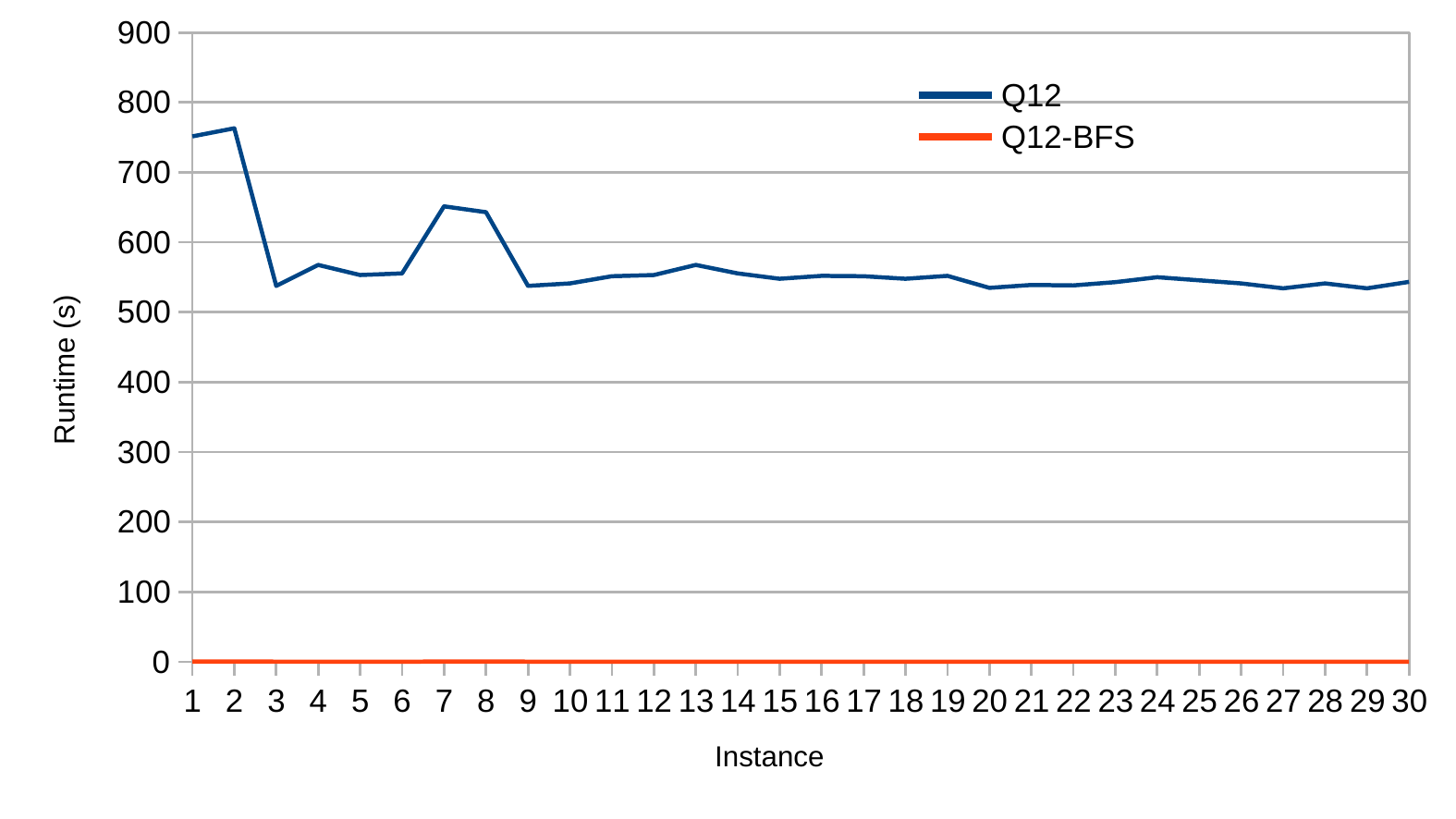}
	\caption{Results for query 12 \textsc{Query12} (average runtime 567.44 seconds) and the optimization approach 1 \textsc{Graph-BFS} (average runtime 0.14 seconds). The speedup factor is 3838.77.}
	\label{img:q12}
\end{figure*} 

Both queries 4 and 12 are pathfinding problems. To retrieve the shortest path, we suggested the execution of a Cypher query using the build in \texttt{shortestPath} algorithm. Applying optimization strategy 1, we suggest the usage of a BFS-approach called \textsc{Graph-BFS}. 

Contrary to expectations, build in Dijkstra's algorithm performs very poor. The runtime lay between 40 and 60 minutes, the average runtime was 2390.44 seconds. In contrast the BFS-approach had a runtime of 1-2 seconds, the average runtime was 1.65 seconds. This is a speedup factor of 1453, see figure \ref{img:q4}.

These results could also be reproduced on Query 12, see figure \ref{img:q12}. The average runtime of \texttt{shortestPath} is 567.44 seconds, approximating 10 minutes. The average runtime of the BFS-approach is 0.14 seconds. This is a speedup factor of 3838.77, see figure \ref{img:q12}. 

These results highlighted that the \texttt{shortestPath} function cannot be used for large scale knowledge graphs due to the runtime. Unexpectedly, the simple BFS-approach utilizing our first optimization strategy decreases the runtime nearly by the factor 3840. Further analysis showed that the speedup is highly influenced by node degree. Nevertheless, \texttt{shortestPath} is unacceptable for information systems with a user frontend.  











\subsection{CRPQ}

\begin{figure*}[t] 
	\centering
	\includegraphics[width=0.95\textwidth]{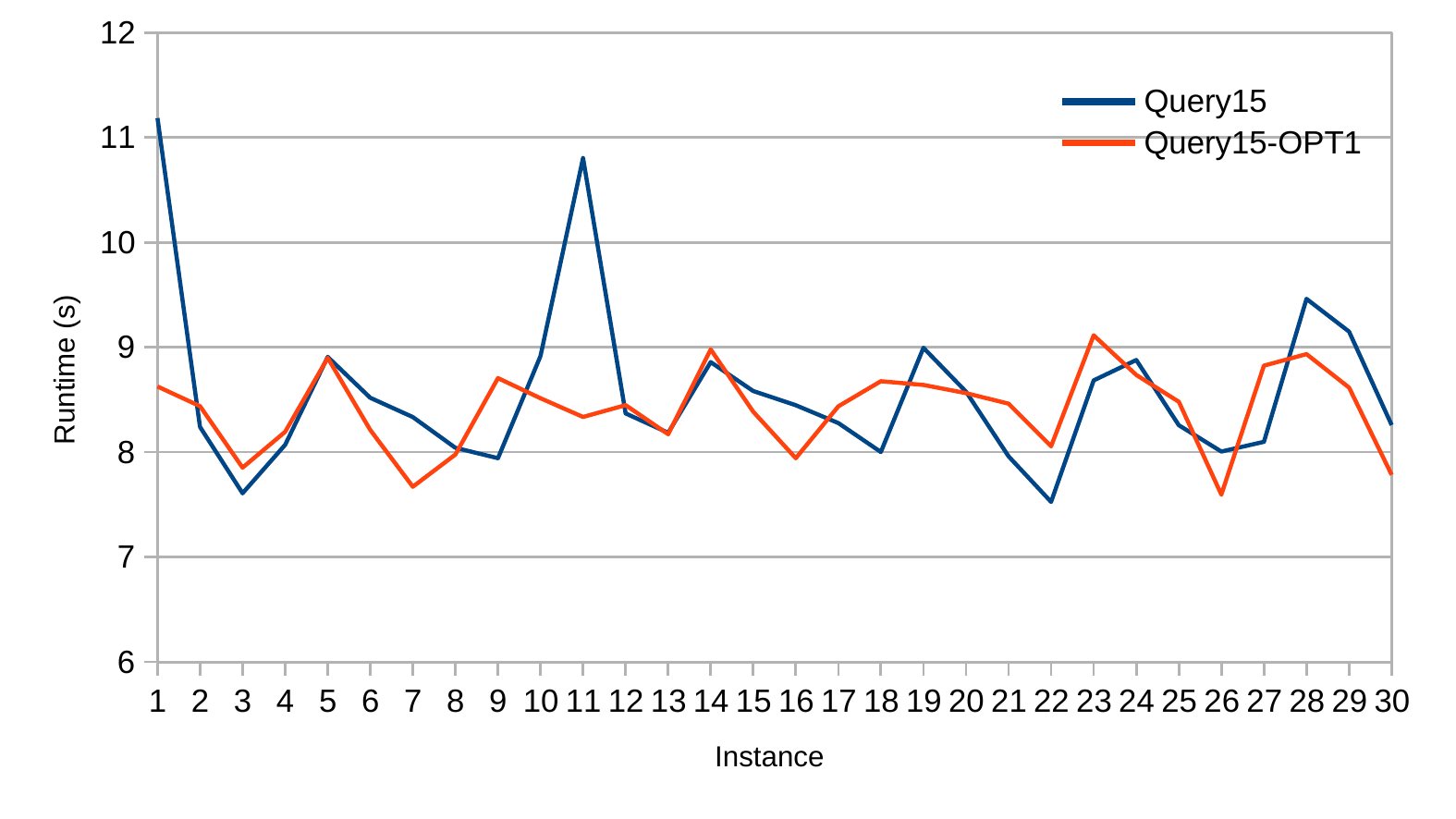}
	\caption{Results for query 15 \textsc{Query15} (average runtime 8.6 seconds) and the optimization approach 1 \textsc{Query15-opt1} (average runtime 8.4 seconds).}
	\label{img:q15}
\end{figure*}

We had a more simple query (15) and a more complex query (1). Regarding \textsc{Query15}, we could only implement our first optimization approach \textsc{Query15-opt1}. Figure \ref{img:q15} presents the runtime data. The average runtime of \textsc{Query15} is 8.6 seconds, the average runtime of \textsc{Query15-opt1} is 8.4 seconds. As we can see, there is no real advantage in applying the optimization approach here. In general both heuristics are competitive, while the simple Cypher query has some situations where it is significantly slower. Although no significant differences were found, the optimization approach shows a rather constant runtime.

\begin{figure*}[t] 
	\centering
	\includegraphics[width=0.95\textwidth]{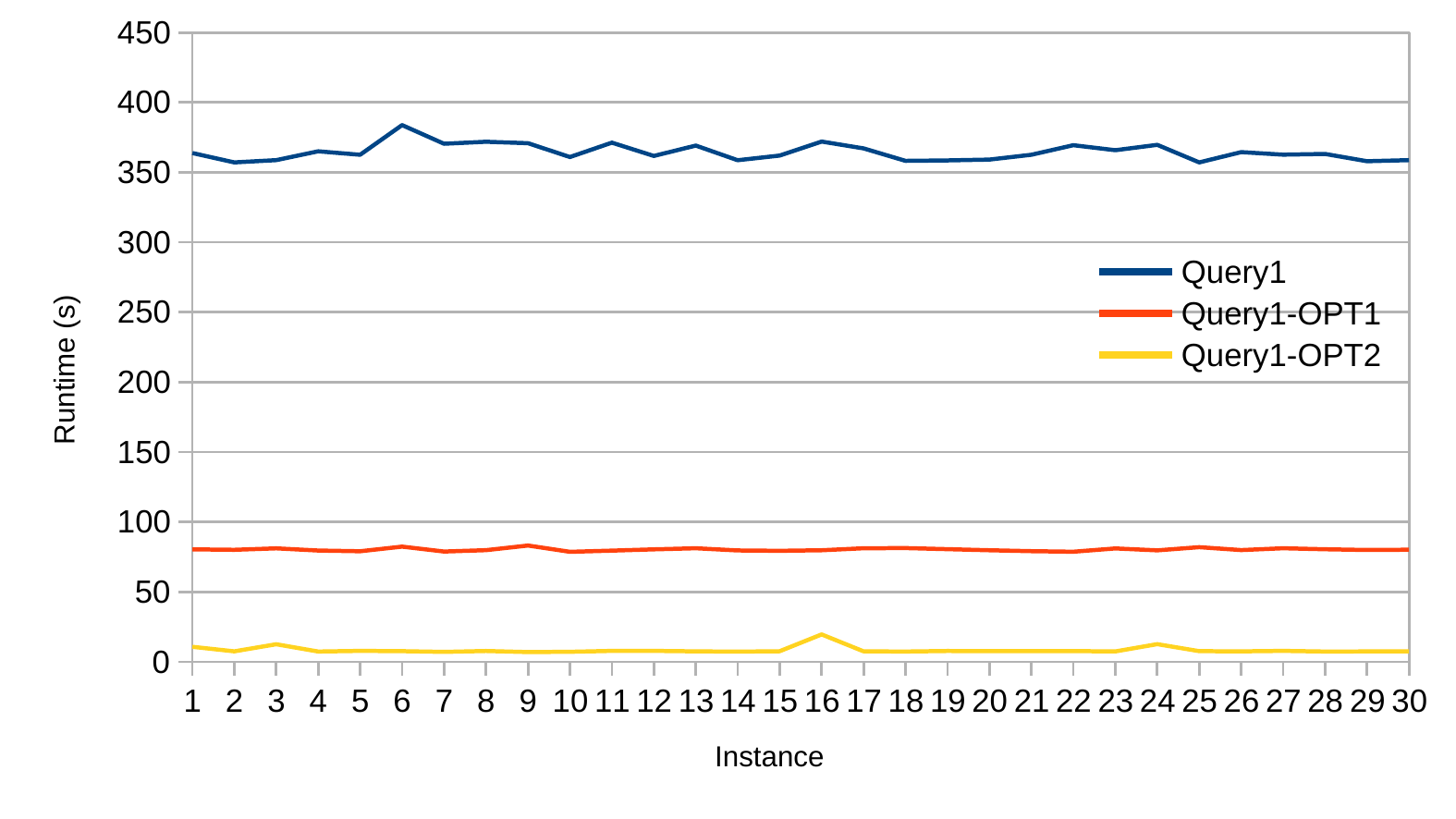}
	\caption{Results for query 1 \textsc{Query1} (average runtime 364.45 seconds)  and the optimization approaches 1 \textsc{Query1-opt1} (average runtime  80.2 seconds)  and 2 \textsc{Query1-opt2} (average runtime  9.6 seconds). In total the speedup factor is 43.8. }
	\label{img:q1}
\end{figure*} 

The most striking results are obtained with more complex queries. The situation changes significantly when analyzing query 1. Here, the Cypher query \textsc{Query1} usually has an execution time of about 7 or 8 minutes, the average runtime is 364.45 seconds. Using the optimization approach 1, the execution time of \textsc{Query1-opt1} reduces to 1-2 minutes, the average runtime is 80.2 seconds. Thus, the runtime decreases by the factor 4.5. Using an polyglot persistence approach and querying SCAIView for the metadata, the execution time of \textsc{Query1-opt2} once again decreases to more or less 10 seconds, in average 9.6 seconds. Here, the runtime decreases by the factor 9,6 compared with \textsc{Query1-opt1} and by the factor 43.8 compared with \textsc{Query1}, see figure {img:q1}. 

It is important to note, that simple queries like Q15 cannot be improved very easy. Graph databases are highly optimized to retrieve relations. But our technique shows a clear advantage over simple Cypher queries when multiple relations are queried, functions for sorting or other purposes are called and especially when single nodes or edges are called to retrieve metadata.  Neo4j shows no good performance when used as a key-value store.

\section{Conclusion and Outlook}

In this paper we presented two new approaches for query optimization on large scale knowledge graphs using graph databases. Knowledge graphs have been shown to play an important role in recent knowledge mining and discovery.  A \emph{knowledge graph} (sometimes also called a \emph{semantic network}) is a systematic way to connect information and data to knowledge on a more abstract level compared to language graphs. 

We used  three  approaches to compare our optimization strategies to state-of-the-art Cypher queries.  Our  goal  was  to  reach the best optimization level without changing the underlying graph database. We believe this solution will aid researchers without a technological background to effectively improve their queries.

Our experiments showed that the proposed optimization strategies can effectively improve the  performance by excluding those parts of queries with the highest runtime. 
Especially the retrieval of single entities like nodes and edges, but also the usage of functions like sorting or shortest paths have been detected for decreasing the execution time significantly. 

Graph databases are highly efficient and optimized for storing and retrieving relations between data points. Thus, we propose to review graph queries carefully and check, if heuristics can be used to merge those parts of a query that are very fast in graph databases. Thus it is an important step to provide a deeper understanding of the underlying graph structures. We could show that most graph queries categorized as locale structures need cannot be executed efficiently out of the box: graph navigation and pattern matching. Only adjacency queries seem to perform very good.   

Although this is a good step towards a better understanding of the underlying problem field, it does not help to find a general solutions to optimize graph queries. Improving the runtime of graph queries needs a careful understanding and improving of the heuristics. 

Our future work includes optimization approaches for federated queries on multiple data sources and better understanding of those cases, where optimization approaches are feasible and lead to a significant improvement of execution time. In addition we plan to evaluate our results with other graph databases like OrientDB.

\section{Acknowledgments}

We thank Marc Jacobs for valuable suggestions, Bruce Schultz for his technical help and both for carefully revising the manuscript. 

This manuscript has been supported by Fraunhofer Society under the MAVO Project; Human Brain Pharmacome.

\bibliographystyle{IEEEtran}
\bibliography{lit}

\begin{thebibliography}{10}
\providecommand{\url}[1]{#1}
\csname url@samestyle\endcsname
\providecommand{\newblock}{\relax}
\providecommand{\bibinfo}[2]{#2}
\providecommand{\BIBentrySTDinterwordspacing}{\spaceskip=0pt\relax}
\providecommand{\BIBentryALTinterwordstretchfactor}{4}
\providecommand{\BIBentryALTinterwordspacing}{\spaceskip=\fontdimen2\font plus
\BIBentryALTinterwordstretchfactor\fontdimen3\font minus
  \fontdimen4\font\relax}
\providecommand{\BIBforeignlanguage}[2]{{%
\expandafter\ifx\csname l@#1\endcsname\relax
\typeout{** WARNING: IEEEtran.bst: No hyphenation pattern has been}%
\typeout{** loaded for the language `#1'. Using the pattern for}%
\typeout{** the default language instead.}%
\else
\language=\csname l@#1\endcsname
\fi
#2}}
\providecommand{\BIBdecl}{\relax}
\BIBdecl

\bibitem{desai2018issues}
M.~Desai, R.~G~Mehta, and D.~P~Rana, ``Issues and challenges in big graph
  modelling for smart city: An extensive survey,'' \emph{International Journal
  of Computational Intelligence \& IoT}, vol.~1, no.~1, 2018.

\bibitem{dorpinghaus2019knowledge}
J.~D{\"o}rpinghaus and A.~Stefan, ``Knowledge extraction and applications
  utilizing context data in knowledge graphs,'' in \emph{2019 Federated
  Conference on Computer Science and Information Systems (FedCSIS)}.\hskip 1em
  plus 0.5em minus 0.4em\relax IEEE, 2019, pp. 265--272.

\bibitem{johnson1977efficient}
D.~B. Johnson, ``Efficient algorithms for shortest paths in sparse networks,''
  \emph{Journal of the ACM (JACM)}, vol.~24, no.~1, pp. 1--13, 1977.

\bibitem{huang2013research}
H.~Huang and Z.~Dong, ``Research on architecture and query performance based on
  distributed graph database neo4j,'' in \emph{2013 3rd International
  Conference on Consumer Electronics, Communications and Networks}.\hskip 1em
  plus 0.5em minus 0.4em\relax IEEE, 2013, pp. 533--536.

\bibitem{holsch2016algebra}
J.~H{\"o}lsch and M.~Grossniklaus, ``An algebra and equivalences to transform
  graph patterns in neo4j,'' in \emph{EDBT/ICDT 2016 Workshops: EDBT Workshop
  on Querying Graph Structured Data (GraphQ)}, 2016.

\bibitem{thakkar2017towards}
H.~Thakkar, D.~Punjani, S.~Auer, and M.-E. Vidal, ``Towards an integrated graph
  algebra for graph pattern matching with gremlin,'' in \emph{International
  Conference on Database and Expert Systems Applications}.\hskip 1em plus 0.5em
  minus 0.4em\relax Springer, 2017, pp. 81--91.

\bibitem{angles2019rdf}
R.~Angles, H.~Thakkar, and D.~Tomaszuk, ``Rdf and property graphs
  interoperability: Status and issues,'' in \emph{Proceedings of the 13th
  Alberto Mendelzon International Workshop on Foundations of Data Management,
  Asunci{\'o}n, Paraguay}, 2019.

\bibitem{mennicke2019modal}
S.~Mennicke, ``Modal schema graphs for graph databases,'' in
  \emph{International Conference on Conceptual Modeling}.\hskip 1em plus 0.5em
  minus 0.4em\relax Springer, 2019, pp. 498--512.

\bibitem{zhao2010graph}
P.~Zhao and J.~Han, ``On graph query optimization in large networks,''
  \emph{Proceedings of the VLDB Endowment}, vol.~3, no. 1-2, pp. 340--351,
  2010.

\bibitem{eymertoward}
J.~Eymer, P.~Dexter, and Y.~D. Liu, ``Toward lazy evaluation in a graph
  database,'' \emph{SPLASH 2019}.

\bibitem{cheung2016sloth}
A.~Cheung, S.~Madden, and A.~Solar-Lezama, ``Sloth: Being lazy is a virtue
  (when issuing database queries),'' \emph{ACM Transactions on Database Systems
  (ToDS)}, vol.~41, no.~2, p.~8, 2016.

\bibitem{mathew2018efficient}
A.~B. Mathew, ``Efficient query retrieval from social data in neo4j using
  lindex.'' \emph{KSII Transactions on Internet \& Information Systems},
  vol.~12, no.~5, 2018.

\bibitem{cabrera2017scalable}
W.~Cabrera and C.~Ordonez, ``Scalable parallel graph algorithms with
  matrix--vector multiplication evaluated with queries,'' \emph{Distributed and
  Parallel Databases}, vol.~35, no. 3-4, pp. 335--362, 2017.

\bibitem{wu2018research}
X.~Wu and S.~Deng, ``Research on optimizing strategy of database-oriented gis
  graph database query,'' in \emph{2018 5th IEEE International Conference on
  Cloud Computing and Intelligence Systems (CCIS)}.\hskip 1em plus 0.5em minus
  0.4em\relax IEEE, 2018, pp. 305--309.

\bibitem{yun2019tkg}
P.~Fournier-Viger, C.~Cheng, L.~J. chuan Wei, U.~Yun, and R.~U. Kiran, ``Tkg:
  Efficient mining of top-k frequent subgraphs,'' in \emph{Big Data Analytics:
  7th International Conference, BDA 2019, Ahmedabad, India, December 17--20,
  2019, Proceedings}, vol. 11932.\hskip 1em plus 0.5em minus 0.4em\relax
  Springer Nature, 2019, p. 209.

\bibitem{fluck}
J.~Fluck, A.~Klenner, S.~Madan, S.~Ansari, T.~Bobic, J.~Hoeng,
  M.~Hofmann-Apitius, and M.~Peitsch, ``Bel networks derived from qualitative
  translations of bionlp shared task annotations,'' in \emph{Proceedings of the
  2013 Workshop on Biomedical Natural Language Processing}, 2013, pp. 80--88.

\bibitem{GO}
M.~Ashburner, C.~A. Ball, J.~A. Blake, D.~Botstein, H.~Butler, J.~M. Cherry,
  A.~P. Davis, K.~Dolinski, S.~S. Dwight, J.~T. Eppig \emph{et~al.}, ``Gene
  ontology: tool for the unification of biology,'' \emph{Nature genetics},
  vol.~25, no.~1, p.~25, 2000.

\bibitem{wishart2017drugbank}
D.~S. Wishart, Y.~D. Feunang, A.~C. Guo, E.~J. Lo, A.~Marcu, J.~R. Grant,
  T.~Sajed, D.~Johnson, C.~Li, Z.~Sayeeda \emph{et~al.}, ``Drugbank 5.0: a
  major update to the drugbank database for 2018,'' \emph{Nucleic acids
  research}, vol.~46, no.~D1, pp. D1074--D1082, 2017.

\bibitem{khan2019consensus}
K.~Khan, E.~Benfenati, and K.~Roy, ``Consensus qsar modeling of toxicity of
  pharmaceuticals to different aquatic organisms: Ranking and prioritization of
  the drugbank database compounds,'' \emph{Ecotoxicology and environmental
  safety}, vol. 168, pp. 287--297, 2019.

\bibitem{huba}
\BIBentryALTinterwordspacing
J.~Dörpinghaus, A.~Stefan, B.~Schultz, and M.~Jacobs. (2020) Towards context
  in large scale biomedical knowledge graphs. [Online]. Available:
  \url{http://arxiv.org/abs/2001.08392}
\BIBentrySTDinterwordspacing

\bibitem{article}
P.~{Barcel{\'{o}} Baeza}, ``{Querying graph databases},'' \emph{Proceedings of
  the ACM SIGACT-SIGMOD-SIGART Symposium on Principles of Database Systems},
  2013.

\bibitem{6313676}
R.~Angles, ``{A Comparison of Current Graph Database Models},'' in \emph{2012
  IEEE 28th International Conference on Data Engineering Workshops}, apr 2012,
  pp. 171--177.

\bibitem{Angles:2017:FMQ:3145473.3104031}
\BIBentryALTinterwordspacing
R.~Angles, M.~Arenas, P.~Barcel{\'{o}}, A.~Hogan, J.~Reutter, and
  D.~Vrgo{\v{c}}, ``{Foundations of Modern Query Languages for Graph
  Databases},'' \emph{ACM Comput. Surv.}, vol.~50, no.~5, pp. 68:1----68:40,
  sep 2017. [Online]. Available: \url{http://doi.acm.org/10.1145/3104031}
\BIBentrySTDinterwordspacing

\bibitem{Wood:2012:QLG:2206869.2206879}
\BIBentryALTinterwordspacing
P.~T. Wood, ``{Query Languages for Graph Databases},'' \emph{SIGMOD Rec.},
  vol.~41, no.~1, pp. 50--60, apr 2012. [Online]. Available:
  \url{http://doi.acm.org/10.1145/2206869.2206879}
\BIBentrySTDinterwordspacing

\bibitem{Pokorny2018}
\BIBentryALTinterwordspacing
J.~Pokorn{\'{y}}, ``{Functional querying in graph databases},'' \emph{Vietnam
  Journal of Computer Science}, vol.~5, no.~2, pp. 95--105, 2018. [Online].
  Available: \url{https://doi.org/10.1007/s40595-017-0104-6}
\BIBentrySTDinterwordspacing

\bibitem{10.1007/978-3-319-24369-6_5}
J.~Pokorny, ``{Graph Databases: Their Power and Limitations},'' in
  \emph{Computer Information Systems and Industrial Management}, K.~Saeed and
  W.~Homenda, Eds.\hskip 1em plus 0.5em minus 0.4em\relax Cham: Springer
  International Publishing, 2015, pp. 58--69.

\bibitem{Needham2019}
M.~Needham and A.~E. Hodler, \emph{{Graph Algorithms}}.\hskip 1em plus 0.5em
  minus 0.4em\relax O'Reilly Media, Inc., 2019.

\bibitem{aziz2010algorithms}
A.~Aziz and A.~Prakash, \emph{Algorithms for Interviews: A Problem Solving
  Approach}.\hskip 1em plus 0.5em minus 0.4em\relax
  algorithmsforinterviews.com, 2010.

\end{thebibliography}

\end{document}